\documentclass[traditabstract]{aa}
\usepackage{graphicx}
\usepackage{txfonts}
 \usepackage{natbib}
 \bibpunct{(}{)}{;}{a}{}{,} 
\newenvironment{system}%
  {\left\lbrace\begin{array}{@{}l@{}}}%
  {\end{array}\right.}

\newcommand{\tc}{\tau_{\rm c}}

\newcommand{\taul}{\tau_\mathrm{c,LEGACY}}

\def\be{\begin{equation}}
\def\ee{\end{equation}}    
\def\ba{\begin{eqnarray}}
\def\ea{\end{eqnarray}}

\def\cnv{v_{\rm T}}
\def\cv{\bar{v}_{\rm T}}
\usepackage{color}

\usepackage{epstopdf}

\begin{document}

\title{A calibration of the Rossby number from asteroseismology}
\author{E. Corsaro\inst{1}\and
A. Bonanno \inst{1}\and
S. Mathur \inst{2,3} \and
R.~A. Garc\'{i}a \inst{4} \and
A.~R.~G. Santos \inst{5} \and
S.~N. Breton\inst{4} \and
{A.~Khalatyan \inst{6}} 
          }
\offprints{Enrico Corsaro\\ \email{enrico.corsaro@inaf.it}}

\institute{
    INAF -- Osservatorio Astrofisico di Catania, via S. Sofia, 78, 95123 Catania, Italy
\and
    Instituto de Astrofísica de Canarias, Santa Cruz de Tenerife, Spain
\and
    Department of Astrophysics, Universidad de La Laguna, Santa Cruz de Tenerife, Spain
\and   
AIM, CEA, CNRS, Université Paris-Saclay, Université Paris Diderot, Sorbonne Paris Cité, F-91191 Gif-sur-Yvette, France
\and 
Department of Physics, University of Warwick, Coventry, CV4 7AL, UK
\and
{Leibniz-Institut für Astrophysik Potsdam (AIP), An der Sternwarte 16, 14482, Potsdam, Germany }
}
   \date{Received 27-05-2021; accepted 15-07-2021}

\abstract{Stellar activity and rotation are tightly related in a dynamo process. Our understanding of this mechanism is mainly limited by our capability of inferring the properties of stellar turbulent convection. In particular, the convective turnover time is a key ingredient through the estimation of the stellar Rossby number, which is the ratio of the rotation period and the convective turnover time. In this work we propose a new calibration of the $(B-V)$ color index dependence of the convective turnover time, hence of the stellar Rossby number. Our new calibration is based on the stellar structure properties inferred through the detailed modeling of solar-like pulsators using asteroseismic observables. We show the impact of this calibration in a stellar activity -- Rossby number diagram by applying it to a sample of about 40,000 stars observed with \textit{Kepler} and for which photometric activity proxy $S_\mathrm{\!ph}$ and surface rotation periods are available. {Additionally, we provide a new calibration of the convective turnover time as function of the $(G_\mathrm{BP}-G_\mathrm{RP})$ color index for allowing applicability in the ESA Gaia photometric passbands.}
}
	  
\titlerunning{}
      \authorrunning{}

\maketitle

\section{Introduction}
\label{sec:intro}
The study of the relationship between stellar activity and  rotation represents one of the  most important tests of stellar dynamo theory. 
In fact, although in solar-like stars various types of dynamo action can be excited
(interface dynamo, $\alpha^2\Omega$ or flux-transport dynamo, \citealt{Brun17review}), 
the dependence of the $\alpha$-effect from basic stellar parameters is not
entirely unconstrained. 
In particular, in a turbulence with length scale $\ell_{\rm T}$, density $\rho$, and rotation $\Omega$, it is possible to show that 
$\alpha\approx \Omega \ell_{\rm T} \nabla (\ln \rho \cnv^2)\approx\epsilon D \Omega$
where $D$ is the density length-scale, $\epsilon=\ell_{\rm T}^2/D^2$, 
and $\cnv$ the convective velocity \citep{1993A&A...269..581R}.
This relation allows us to express the dynamo number $C_\alpha$ in terms of the rotation rate as 
$C_\alpha ={D\alpha}/{\eta}=3 \tau \Omega$
where $\eta$ is the turbulent (eddies) diffusivity and  $\tau$ is the convective
turnover time. 
It is thus apparent that a larger value of $\tau \Omega$ (often called Coriolis number) 
implies a lower threshold for the onset 
of the dynamo. However, while the rotation is in principle 
an observable quantity, in general $\tau$ depends on the efficiency of convection
and therefore on the individual fundamental stellar parameters. 

While interpreting observational data from low-mass stars it is customary to weigh
the relative importance of turbulent convection against rotation in terms of the 
stellar Rossby number as follows:
\begin{equation}
\label{rossby}
    Ro=\frac{P_{\rm rot}}{\tau} \, ,
\end{equation}
where $P_{\rm rot}$ is the surface rotation period, while $\tau$ is the convective turnover time given as
\begin{equation}
\label{tau}
\tau=\frac{d_{\rm CZ}}{\cv}\, ,
\end{equation}
with $d_{\rm CZ}$ being the thickness of the convective
envelope and $\cv $ the average convective velocity \citep{Brun17}.

In practice $\tau$ is often determined from the semi-empirical $(B-V)$ flux excess
dependence obtained by \cite{Noyes84Rossby}, where  a fixed value of the mixing-length 
parameter has been used.  The limitation of this approach is that 
already in the case of the Sun, 
it yields $\tau \sim 12$ days, a value significantly smaller
than the one obtained from a standard solar model, i.e. $\sim 45$ days \citep{Bonanno02}.

If the turbulence is determined by only one length scale, it is not
difficult to evaluate Eq.~(\ref{tau}) from the fundamental stellar astrophysical parameters. 
In this case, if $T$ is the temperature,  
$c_p$ the specific heat at constant pressure and $H_{\rm p}$ the pressure scale height, 
the convective fluxes can be expressed as
\begin{eqnarray}
&&    F_{\rm conv}=\rho c_p T 
    \left ( \frac{\ell_{\rm T}}{H_{\rm p}}\right)^2\sqrt{\frac{1}{2}g H_p} 
    (\nabla-\nabla_{ad})^{3/2} \nonumber \\
&& \approx \frac{15}{8\pi \sqrt{2}} \frac{M}{R^3} \left (\frac{\ell_{\rm T}}{H_{\rm p}}\right)^2 
    \left(\frac{G M}{R}\right)^{3/2}(\nabla-\nabla_{ad})^{3/2} \nonumber\\
&& \approx \frac{M}{R^3}\left(\frac{G M}{R}\right)^{3/2}(\nabla-\nabla_{ad})^{3/2}
\end{eqnarray}
where in the last line we assumed $\ell_{\rm T}/H_{\rm p} \approx 1.6$, a typical value for
a solar-like star.
On the other hand, $F_{\rm conv}\approx L/R^2$ and therefore 
\begin{equation}
    (\nabla-\nabla_{ad})\approx \left (\frac{L R}{M}\right)^{2/3} \frac{R}{GM} \, .
\end{equation}
With $v_s$ denoting the sound speed, we thus have
\begin{equation}
\label{cv}
    \cv \approx v_s \sqrt{\nabla-\nabla_{ad}} = \left( \frac{L R}{M}\right)^{1/3} \, ,
\end{equation}
where in the last line we made use 
of the fact that $v_s\approx \sqrt{ g H_{\rm p}} \approx \sqrt{G M/R}$.
It is interesting to notice that the estimation of Eq.~(\ref{cv}) in Eq.~(\ref{tau})
represents indeed a rather good estimate of the {\it local} convective turnover
time at the base of the convection zone obtained from stellar models. 
In the case of the Sun for instance one obtains 
$\tau  \approx 45$ days, a value consistent with 
the local value
of $(H_{\rm p} / \cnv)_{r=0.72 R_\odot}$ for a fully calibrated solar standard model \citep{Bonanno02}, 
in agreement with the calculation of \cite{landin2010}.

While the basic stellar parameters like luminosity, mass, 
and radius in Eq.~(\ref{cv}) can easily be obtained at least for nearby stars, 
a reliable estimation of $d_{\rm CZ}$ is in general a much more difficult task.
Asteroseismology, however, can in principle provide us with this crucial 
piece of information for stars with well characterized oscillation properties \citep[e.g. see][]{Garcia19}. Similarly, taking advantage of asteroseismic modeling, an estimation of the convective turnover time was done for 10 stars by \citet{Mathur14b}. The authors computed the integral of the convective velocity over the convection zone but using the profile of the convective velocity from the best-fit model obtained with seismic observables. 

In this work we thus propose to calibrate the empirical relation by \citet{Noyes84Rossby} by 
exploiting a sample of main-sequence and sub-giant stars observed by the NASA \textit{Kepler} mission \citep{Borucki10,Koch10} and exhibiting solar-like oscillations. 
In particular we shall make use of this new calibration 
in a large sample of \textit{Kepler} solar-like stars with known surface rotation periods, $P_\mathrm{rot}$, and photometric activity indexes, $S_\mathrm{\!ph}$.  Brightness variations due to active regions co-rotating with the stellar surface provide constraints on both stellar properties \citep[e.g.][]{Garcia10Science,Reinhold13,Nielsen13,McQuillan14,Mathur14,Salabert16Sph,Salabert17Sun,Gordon21}. In this work, we show how the photometric activity index $S_\mathrm{\!ph}$ correlates with the Rossby number in a sample of $\sim$\,40,000 stars from \citet[][in press]{Santos19}.

\section{Observations \& data}
Our observational set comprises two samples of stars. The first sample consists of stars for which a detailed asteroseismic analysis is available. This sample is used in Sect.~\ref{sec:bayes} for obtaining a new calibration of the convective turnover time. The second sample contains stars that have a measure of their activity level and surface rotation, and it is used in Sect.~\ref{sec:results} to illustrate the implications of adopting our new calibration.

\subsection{Calibration sample}
To address the limitation of previous works aiming at obtaining a reliable estimate of the convective turnover time \citep[e.g.][]{Noyes84Rossby,Lehtinen21,See21Rossby}, we selected 62 G- and F-type stars from the {\it Kepler} LEGACY 
sample for which a precise determination of stellar parameters
and chemical composition is available
\citep{Lund17LEGACY,Nissen17LEGACY,SilvaAguirre17}, along with $B_T$, $V_T$ magnitudes from the Tycho-2 catalog \citep{Hog2000}. The LEGACY sample consists of the best characterized main-sequence and sub-giant stars observed by {\it Kepler}. Through the use of asteroseismology, the wealth of detailed oscillation mode properties measured in these stars allowed us to probe their internal structure with a high level of detail, hence in particular constraining the position of the base of the convection zone (CZ). This is essentially possible thanks to the probing power of individual mode frequencies, which are found in a large number for these stars thanks to the high-quality \emph{Kepler} photometric observations, spanning more than three years in high duty-cycle. Stellar mass, radius, as well as the position of the base of the CZ have been inferred with high precision, i.e., about 2\% in radius and 4\% in mass. As already described in Sect.~\ref{sec:intro}, these are key quantities needed to define the convective turnover time in these stars.

\subsection{Stellar activity sample}
\label{sec:activity}
The long-term continuous observations collected by {\it Kepler} are suitable to constrain stellar magnetic activity and rotation. Recently, using KEPSEISMIC\footnote{https://archive.stsci.edu/prepds/kepseismic/} light curves \citep{Garcia11,Garcia14,Pires15} of mid-F to M stars observed by {\it Kepler}, the sample of known rotation periods, $P_{\rm rot}$, and photometric activity proxies, $S_{\rm\! ph}$, was greatly extended \citep[][Santos et al. in press]{Santos19}. Indeed, more than 60\% new $P_{\rm rot}$ detections were reported compared to the previous largest rotation-period catalog \citep{McQuillan14}. The rotation-period candidates were obtained by combining the wavelet analysis and the autocorrelation function of light curves \citep{Mathur10,Garcia14activity,Ceillier16,Ceillier17}, while the reliable rotation periods were selected through a machine learning algorithm (ROOSTER; \citealt{Breton21}), automatic selection, and visual inspection \citep[][Santos et al. in press]{Santos19}. Having $P_{\rm rot}$, $S_{\rm\! ph}$ is computed as the standard deviation over light curve segments of length $5\times P_{\rm rot}$ \citep{Mathur14b}. \cite{Salabert16Sph,Salabert17Sun} have shown that $S_{\rm\! ph}$ is an appropriate proxy for stellar magnetic activity.

Starting with a sample of 55,232 stars from \citet[][]{Santos19}, Santos et al. (in press), Mathur et al. (in prep.), including F, G, K, and M dwarfs as well as subgiants, we first removed the subgiants by making a crude cut on the surface gravity taking only stars with $\log g$ above 4.2\,dex. We then collected the $B$ and $V$ magnitudes from the American  Association  of  Variable  Star  Observers  Photometric  All-Sky  Survey \citep{Henden15}\footnote{https://cdsarc.unistra.fr/viz-bin/cat/II/336}. We further discarded stars flagged as close-in binary candidates by \citet{Santos19,Santos21}, which led us to a sample of 39,125 stars with inferred $P_{\rm rot}$ and $S_{\rm\!ph}$. 

\begin{figure*}[t]
   \centering
   \includegraphics[width=1.0\textwidth]{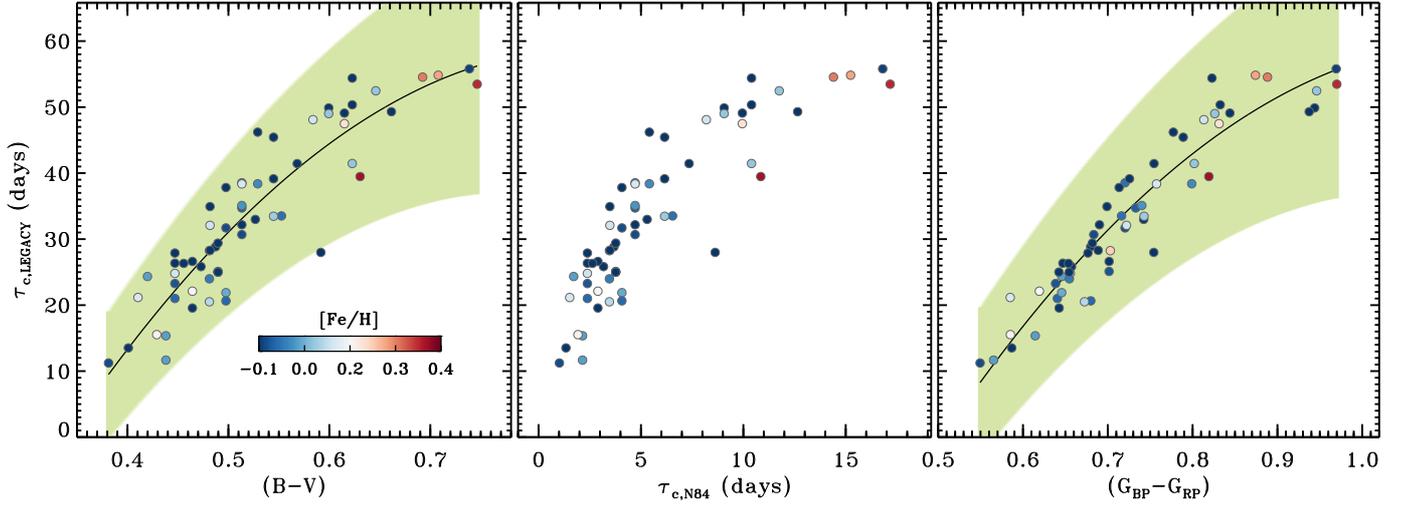}
\caption{Convective turnover time for the LEGACY sample, $\tau_\mathrm{c,LEGACY}$. \textit{Left panel:} Quadratic fit using Eq.~(\ref{eq:tau_quadratic}) as a function of $(B-V)$ color as a solid black line, with the green shaded area representing the $1$-$\sigma$ credible region obtained from the Bayesian inference. The color coding shows the stellar metallicity reported by \citealt{Lund17LEGACY}. \textit{Right panel:} Similar as for the left panel, but showing the result as a function of the Gaia $(G_\mathrm{BP}-G_\mathrm{RP})$ color, with the quadratic fit from Eq.~(\ref{eq:tau_gaia_quadratic}) overlaid. \textit{Middle panel:} Comparison between $\tau_\mathrm{c,LEGACY}$ and $\tau_\mathrm{c,N84}$ as obtained by means of the $(B-V)$ color.}
         \label{fig:noyes_legacy}
\end{figure*}

\section{Bayesian inference}
\label{sec:bayes}
Following from Eq.~(\ref{cv}) and Eq.~(\ref{tau}) $\tc$ can be expressed in terms of fundamental stellar properties as
\begin{equation}
\tc \simeq (R-R_\mathrm{CZ}) \left( \frac{M}{L R} \right)^{1/3} \, ,
\label{eq:tau_def}
\end{equation} 
where in the case of the LEGACY sample the quantities $R$, $R_\mathrm{CZ}$, $L$, $M$ are obtained from the stellar and asteroseismic modeling performed by \cite{SilvaAguirre17} \footnote{http://cdsarc.unistra.fr/viz-bin/nph-Cat/html?J/ApJ/835/173/table4.dat} using the ASTFIT pipeline, a combination of ASTEC (Aarhus STellar Evolution Code) for stellar evolutionary models \citep{CD08ASTEC} and ADIPLS (Aarhus adiabatic oscillation package) for theoretical pulsation frequencies \citep{CD08ADIPLS}. We therefore compare the outcome of Eq.~(\ref{eq:tau_def}), hereafter $\tau_\mathrm{c,LEGACY}$, to that obtained from the original semi-empirical relation presented by \cite{Noyes84Rossby}, namely
\begin{equation}
\log \tau_c = 
\begin{system}
1.362 - 0.166x + 0.025x^2 -5.323x^2 \qquad x > 0 \\
1.362 -0.14x \qquad x < 0 \, , \\
\end{system}
\end{equation}
where $x = 1 - (B-V)$, and which we refer to as $\tau_\mathrm{c,N84}$ for clarity.
We note that the relation presented by \cite{Noyes84Rossby} has to be evaluated by means of the Johnson-Cousin $B$ and $V$ magnitudes, which differ from those available for the LEGACY sample, corresponding to Tycho catalog $B_T$ and $V_T$ magnitudes. For obtaining a proper comparison between the two convective turnover times here considered we therefore apply a transformation converting our Tycho $(B_T-V_T)$ color index into a $(B-V)$ one following ESA97 Vol. 1, Sect. 1.3.

\subsection{Calibration with $B-V$ color}
Our basic assumption is that a relation between $\tau_\mathrm{c,LEGACY}$ and $(B-V)$ color could be represented by a linear law of the type
\begin{equation}
\label{eq:tau_linear}
\tau_\mathrm{c,LEGACY} = a_1 + a_2 (B-V)\quad\mbox{(days)} \, ,
\end{equation}
However, we also incorporate the possibility of having departures from a simple linear relation. We therefore consider an additional law involving a quadratic dependence on $(B-V)$ of the type
\begin{equation}
\label{eq:tau_quadratic}
\tau_\mathrm{c,LEGACY} = b_1 + b_2 (B-V) + b_3 (B-V)^2 \quad\mbox{(days)} \, .
\end{equation}
For obtaining the estimates of the free parameters $(a_1,a_2)$ of Eq.~(\ref{eq:tau_linear}) and $(b_1,b_2,b_3)$ of Eq.~(\ref{eq:tau_quadratic}) we make use of a Bayesian parameter estimation. This procedure is carried out by means of the public tool \textsc{D\large{iamonds}}\footnote{https://github.com/EnricoCorsaro/DIAMONDS} \citep[high-DImensional And multi-MOdal NesteD Sampling,][]{Corsaro14} based on the nested sampling Monte Carlo algorithm \citep{Skilling04}. Here we adopt a standard Normal likelihood function 
assuming a conservative theoretical uncertainty of 1 day on each estimate of $\taul$, and uniform priors on all free parameters.

For testing the significance of having a quadratic dependence between $\taul$ and $(B-V)$, we perform a Bayesian model comparison by means of the Bayesian evidence computed by \textsc{D\large{iamonds}}. In this case we compare the model obtained using Eq.~(\ref{eq:tau_linear}) with the one of Eq.~(\ref{eq:tau_quadratic}). We find that the corresponding Bayes' factor $\mathcal{B}_\mathrm{quadratic,linear} \simeq 95$ significantly exceeds the condition for a strong evidence \citep{Jeffreys61}, hence strongly favoring the scenario of a quadratic dependence. This implies that the adoption of Eq.~(\ref{eq:tau_quadratic}) is by far statistically justified when adopting the LEGACY sample as a calibration set. The result of this inference process is presented in Fig.~\ref{fig:noyes_legacy} (left panel), where the quadratic trend appears dominant along the entire range investigated ($0.38 < (B-V) < 0.75$, corresponding to $1 \lesssim \tau_\mathrm{c,N84} \lesssim 17$\,days), with the coefficients $b_1 = -101.2^{+5.8}_{-4.7}$\,d, $b_2 = 373.4^{+18.5}_{-21.6}$\,d mag$^{-1}$, and $b_3 = -217.6^{+19.0}_{-17.3}$\,d mag$^{-2}$. We further point out that this relation is intrinsically incorporating the effect of a varying metallicity from star to star because the stellar properties used for the evaluation of the convective turnover time are the product of a detailed asteroseismic and stellar structure modeling that also take into account metallicity. This can be seen in Fig.~\ref{fig:noyes_legacy}, where the color-coding of the symbols indicating the metallicities adopted by \cite{Lund17LEGACY} shows no evidence of a preferential trend with $\tau_\mathrm{c,LEGACY}$.

\begin{figure*}[ht]
   \centering
   \includegraphics[width=1.0\textwidth]{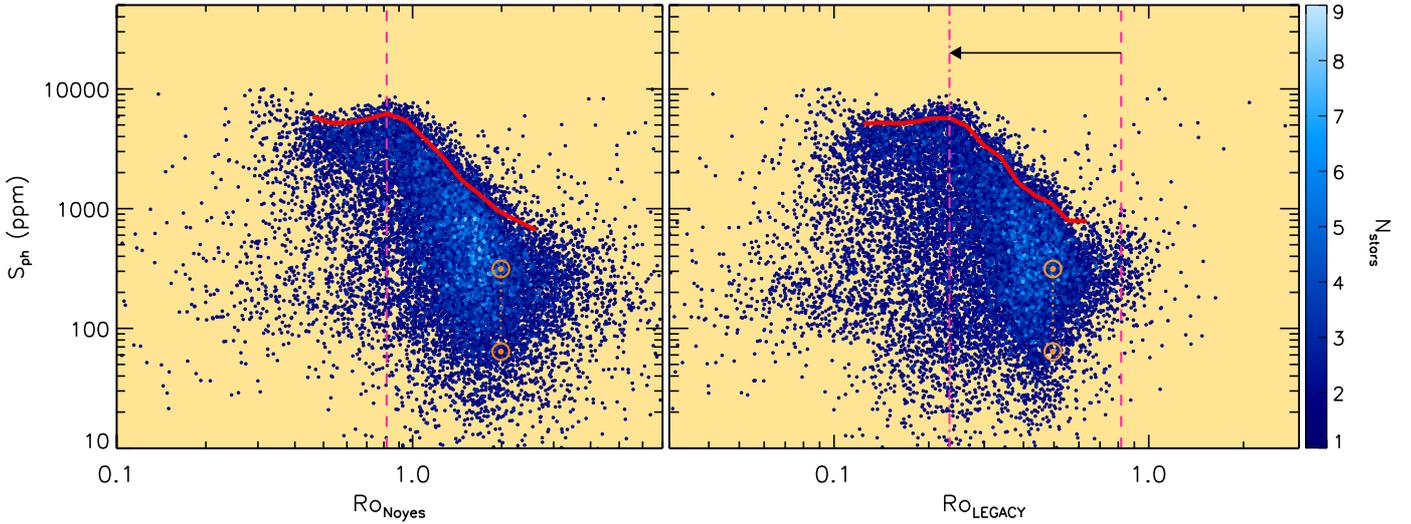}
\caption{Stellar activity index $S_\mathrm{\!ph}$ as a function of the Rossby number. \textit{Left panel}: Rossby number evaluated with the semi-empirical relation by \citealt{Noyes84Rossby}. \textit{Right panel}: similar to the left panel but using the Rossby number evaluated with the new convective turnover time from Eq.~(\ref{eq:tau_quadratic}). The thick red lines represent the maximum $S_\mathrm{\!ph}$ of the 95 percentile of each chunk in Rossby number, evaluated in a range centered around the kink. Sun symbols mark the position of the Sun between the minimum and maximum of activity ($Ro_\mathrm{Noyes,\odot} =1.989$, $Ro_\mathrm{LEGACY,\odot} = 0.496$). The position of the kink is marked by the vertical dashed and dot-dashed lines, for the left and right panel, respectively. The left-pointing arrow shows how the position of the kink shifts when switching from $\tau_\mathrm{c,N84}$ to $\tau_\mathrm{c,LEGACY}$. The density of stars is indicated by the blue color scale.}
         \label{fig:ro_legacy}
   \end{figure*}

\subsection{Calibration with Gaia $G_\mathrm{BP}-G_\mathrm{RP}$ color}
\label{sec:gaia}
We adopt new Gaia EDR3 $(G_\mathrm{BP}-G_\mathrm{RP})$ colors \citep{GaiaEDR3} for studying the relation $\taul$ -- $(G_\mathrm{BP}-G_\mathrm{RP})$. This relation is of particular usefulness and applicability for obtaining a reliable estimation of the stellar convective turnover time in the context of the very large catalog parameters available from ESA Gaia. 

Following the approach presented for the calibration of a $\taul$ -- $(B-V)$ relation, we perform a Bayesian parameter estimation for both a linear and a quadratic model of the type
\begin{equation}
\label{eq:tau_gaia_linear}
\tau_\mathrm{c,LEGACY} = a'_1 + a'_2 (G_\mathrm{BP}-G_\mathrm{RP})\quad\mbox{(days)} \, ,
\end{equation}
and 
\begin{equation}
\label{eq:tau_gaia_quadratic}
\tau_\mathrm{c,LEGACY} = b'_1 + b'_2 (G_\mathrm{BP}-G_\mathrm{RP}) + b'_3 (G_\mathrm{BP}-G_\mathrm{RP})^2\quad\mbox{(days)} \, .
\end{equation}
Similarly to what obtained for the Johnson-Cousin photometric system, our Bayesian model comparison process, resulting from a Bayes' factor $\ln \mathcal{B}_\mathrm{linear,quadratic} \simeq 101$, indicates that a quadratic model is strongly favored in interpreting the observed trend when using Gaia colors in the range $0.55 < (G_\mathrm{BP}-G_\mathrm{RP}) < 0.97$. This trend is already evident from Fig.~\ref{fig:noyes_legacy} (right panel). The estimated parameters of the quadratic model are $b'_1 = -134.0^{+6.2}_{-5.9}$\,d, $b'_2 = 341.7^{+16.4}_{-16.2}$\,d mag$^{-1}$, and $b'_3 = -150.6^{+10.8}_{-10.5}$\,d mag$^{-2}$.

\section{Results}
\label{sec:results}
Our Eq.~(\ref{eq:tau_quadratic}), by means of a direct dependence from the $(B-V)$ color index, allows us to estimate a proper turbulent time-scale for a dynamo action deeply seated at the bottom of the CZ, at least for stars close to the main sequence. The estimated convective turnover time for the LEGACY sample appears to be strongly correlated to that obtained from the standard relation by \cite{Noyes84Rossby}, although this dependence is clearly not linear (Fig. 2, middle panel). In addition, we have extended the applicability of this kind of dependence to the Gaia photometric system by calibrating another relation that can be used to obtain a convective turnover time directly from the $(G_\mathrm{BP}-G_\mathrm{RP})$ color index.

To illustrate the implications of our new calibration, we have applied it in comparison to the standard Noyes' formulation \citep{Noyes84Rossby} for a large sample of stars observed by {\it Kepler} with measured rotation periods and presented in Sect.~\ref{sec:activity}. We show the result in Fig.~\ref{fig:ro_legacy}, where the photometric activity index $S_\mathrm{\!ph}$ is plotted as a function of the Rossby number, evaluated using both the Noyes' prescription (left panel of Fig.~\ref{fig:ro_legacy}) and our new calibration (right panel of Fig.~\ref{fig:ro_legacy}) for stars having $(B-V)$ within the range provided for our new calibration (see Sect.~\ref{sec:bayes}, 16844 targets in total). For highlighting the presence of a kink in the distribution we have binned the sample into 50 chunks having constant size in logarithm of Rossby number. For each chunk we computed the maximum out of the lowest 95 percentile in $S_\mathrm{\!ph}$ of the corresponding portion of star sample. The outcome is shown as a thick red line, which clearly follows the shape of the top edge of the distribution. Here it can be seen that the kink corresponding to the change in regime from saturation to a linear dependence of the photometric activity index as a function of the Rossby number shifts from $Ro \sim 0.82$ when using the Noyes' prescription \citep{Noyes84Rossby} to $Ro \sim 0.23$ when adopting our newly calibrated $\tc$ from Eq.~(\ref{eq:tau_quadratic}). The extent of the shift of the kink position in Rossby number \citep[e.g.][]{See21Rossby} is emphasized by the left-pointing arrow in the right panel of Fig.~\ref{fig:ro_legacy}. The position of the Sun is also shown, with $S_\mathrm{\!ph}$ computed at the minimum and maximum of activity \citep{Mathur19activity}. It is reassuring that our result based on the adoption of $\tau_\mathrm{c,LEGACY}$ appears to be in agreement with the findings by \cite{See21Rossby}. In their work, the authors adopted the photometric variability amplitude $R_\mathrm{per}$ originally introduced by \cite{Basri10}, and a Rossby number obtained from model structure grids and rotation periods from \cite{McQuillan14}. 

Finally, we point out that care is needed when dealing with more massive F-type stars ($M \gtrsim 1.3 M_{\odot}$). Because of their thin convective envelopes, these stars happen to fall at the lower edge of the range of convective turnover times covered by our calibration set (say $\tau_\mathrm{N84} < 3\,$days). This implies that the derived quadratic law from Eq.~(\ref{eq:tau_quadratic}) when adopting $(B-V)$, and from Eq.~(\ref{eq:tau_gaia_quadratic}) for the $(G_\mathrm{BP}-G_\mathrm{RP})$ colors, may be less accurate in this regime. Future improvements of the proposed relation will aim at including additional stars in this region of small convective turnover time within the calibration sample.

\section{Conclusions}
The estimation of a proper convective turnover time is essential for our comprehension of the stellar dynamo action through the Rossby number. In this work we have shown that starting from the observed color of the star, and specifically the $(B-V)$ color index from the Johnson-Cousin photometric system and the $(G_\mathrm{BP}-G_\mathrm{RP})$ color index from ESA Gaia, it is possible to obtain a reliable estimate of a local convective turnover time that is comparable to the one obtained by stellar models and evaluated at the bottom of the CZ. In particular, we note that through the adoption of Eq.~(\ref{eq:tau_def}) one can evaluate a proper convective turnover time for large samples of stars without the need of computing stellar models. This is supported by the agreement found in the location of the kink of the relation between our photometric stellar activity index, $S_\mathrm{\!ph}$, and the Rossby number (see Fig.~\ref{fig:ro_legacy}), and the one presented by \cite{See21Rossby}, where the computation of the convective turnover time is instead based on the adoption of stellar model grids.

We consider that these new calibrations pave the way for future studies of the dynamo theory that are based on the adoption of photometric activity indices obtained from past and ongoing space missions such as NASA \textit{Kepler}, K2 \citep{Howell14K2}, the NASA Transiting Exoplanet Survey Satellite (TESS) \citep{Ricker14TESS}, as well as the soon to come ESA PLanetary Transits and Oscillations of stars (PLATO) \citep{Rauer14PLATO}. This result opens the possibility to perform detailed statistical studies that can be extended to very large stellar populations spanning a wide range of fundamental properties and rotation. This is of particular relevance in the context of the vast amount of color information being provided by ESA Gaia space mission. Future improvements on the stellar modeling through the inclusion of internal dynamical processes, e.g. angular momentum transport, could be exploited to further refine the current relationships.

\section*{Acknowledgments}
We thank the referee Dr. Travis Metcalfe for valuable comments that helped in improving the manuscript. E.C. and A.B. acknowledge support from PLATO ASI-INAF agreement no. 2015-019-R.1-2018. S.M. acknowledges support by the Spanish Ministry of Science and Innovation with the Ramon y Cajal fellowship number RYC-2015-17697 and the grant number PID2019-107187GB-I00. R.A.G. and S.N.B. acknowledge the support from the CNES GOLF and PLATO grants. A.R.G.S. acknowledges the support of the STFC consolidated grant ST/T000252/1 and NASA grant No. NNX17AF27.

\bibliographystyle{aa}
\bibliography{tau}

\begin{thebibliography}{46}
\expandafter\ifx\csname natexlab\endcsname\relax\def\natexlab#1{#1}\fi

\bibitem[{{Basri} {et~al.}(2010){Basri}, {Walkowicz}, {Batalha}, {Gilliland},
  {Jenkins}, {Borucki}, {Koch}, {Caldwell}, {Dupree}, {Latham}, {Meibom},
  {Howell}, \& {Brown}}]{Basri10}
{Basri}, G., {Walkowicz}, L.~M., {Batalha}, N., {et~al.} 2010, \apjl, 713, L155

\bibitem[{{Bonanno} {et~al.}(2002){Bonanno}, {Schlattl}, \&
  {Patern{\`o}}}]{Bonanno02}
{Bonanno}, A., {Schlattl}, H., \& {Patern{\`o}}, L. 2002, \aap, 390, 1115

\bibitem[{{Borucki} {et~al.}(2010){Borucki}, {Koch}, {Basri}, {Batalha},
  {Brown}, {Caldwell}, {Caldwell}, {Christensen-Dalsgaard}, {Cochran},
  {DeVore}, {Dunham}, {Dupree}, {Gautier}, {Geary}, {Gilliland}, {Gould},
  {Howell}, {Jenkins}, {Kondo}, {Latham}, {Marcy}, {Meibom}, {Kjeldsen},
  {Lissauer}, {Monet}, {Morrison}, {Sasselov}, {Tarter}, {Boss}, {Brownlee},
  {Owen}, {Buzasi}, {Charbonneau}, {Doyle}, {Fortney}, {Ford}, {Holman},
  {Seager}, {Steffen}, {Welsh}, {Rowe}, {Anderson}, {Buchhave}, {Ciardi},
  {Walkowicz}, {Sherry}, {Horch}, {Isaacson}, {Everett}, {Fischer}, {Torres},
  {Johnson}, {Endl}, {MacQueen}, {Bryson}, {Dotson}, {Haas}, {Kolodziejczak},
  {Van Cleve}, {Chandrasekaran}, {Twicken}, {Quintana}, {Clarke}, {Allen},
  {Li}, {Wu}, {Tenenbaum}, {Verner}, {Bruhweiler}, {Barnes}, \&
  {Prsa}}]{Borucki10}
{Borucki}, W.~J., {Koch}, D., {Basri}, G., {et~al.} 2010, Science, 327, 977

\bibitem[{{Breton} {et~al.}(2021){Breton}, {Santos}, {Bugnet}, {Mathur},
  {Garc{\'\i}a}, \& {Pall{\'e}}}]{Breton21}
{Breton}, S.~N., {Santos}, A.~R.~G., {Bugnet}, L., {et~al.} 2021, \aap, 647,
  A125

\bibitem[{{Brun} \& {Browning}(2017)}]{Brun17review}
{Brun}, A.~S. \& {Browning}, M.~K. 2017, Living Reviews in Solar Physics, 14, 4

\bibitem[{{Brun} {et~al.}(2017){Brun}, {Strugarek}, {Varela}, {Matt},
  {Augustson}, {Emeriau}, {DoCao}, {Brown}, \& {Toomre}}]{Brun17}
{Brun}, A.~S., {Strugarek}, A., {Varela}, J., {et~al.} 2017, \apj, 836, 192

\bibitem[{{Ceillier} {et~al.}(2017){Ceillier}, {Tayar}, {Mathur}, {Salabert},
  {Garc{\'\i}a}, {Stello}, {Pinsonneault}, {van Saders}, {Beck}, \&
  {Bloemen}}]{Ceillier17}
{Ceillier}, T., {Tayar}, J., {Mathur}, S., {et~al.} 2017, \aap, 605, A111

\bibitem[{{Ceillier} {et~al.}(2016){Ceillier}, {van Saders}, {Garc{\'\i}a},
  {Metcalfe}, {Creevey}, {Mathis}, {Mathur}, {Pinsonneault}, {Salabert}, \&
  {Tayar}}]{Ceillier16}
{Ceillier}, T., {van Saders}, J., {Garc{\'\i}a}, R.~A., {et~al.} 2016, \mnras,
  456, 119

\bibitem[{{Christensen-Dalsgaard}(2008{\natexlab{a}})}]{CD08ADIPLS}
{Christensen-Dalsgaard}, J. 2008{\natexlab{a}}, \apss, 316, 113

\bibitem[{{Christensen-Dalsgaard}(2008{\natexlab{b}})}]{CD08ASTEC}
{Christensen-Dalsgaard}, J. 2008{\natexlab{b}}, \apss, 316, 13

\bibitem[{{Corsaro} \& {De Ridder}(2014)}]{Corsaro14}
{Corsaro}, E. \& {De Ridder}, J. 2014, \aap, 571, A71

\bibitem[{{Gaia Collaboration} {et~al.}(2021){Gaia Collaboration}, {Brown},
  {Vallenari}, {Prusti}, {de Bruijne}, {Babusiaux}, {Biermann}, {Creevey},
  {Evans}, {Eyer}, \& et~al.}]{GaiaEDR3}
{Gaia Collaboration}, {Brown}, A.~G.~A., {Vallenari}, A., {et~al.} 2021, \aap,
  649, A1

\bibitem[{{Garc{\'\i}a} \& {Ballot}(2019)}]{Garcia19}
{Garc{\'\i}a}, R.~A. \& {Ballot}, J. 2019, Living Reviews in Solar Physics, 16,
  4

\bibitem[{{Garc{\'\i}a} {et~al.}(2014{\natexlab{a}}){Garc{\'\i}a}, {Ceillier},
  {Salabert}, {Mathur}, {van Saders}, {Pinsonneault}, {Ballot}, {Beck},
  {Bloemen}, {Campante}, {Davies}, {do Nascimento}, {Mathis}, {Metcalfe},
  {Nielsen}, {Su{\'a}rez}, {Chaplin}, {Jim{\'e}nez}, \&
  {Karoff}}]{Garcia14activity}
{Garc{\'\i}a}, R.~A., {Ceillier}, T., {Salabert}, D., {et~al.}
  2014{\natexlab{a}}, \aap, 572, A34

\bibitem[{{Garc{\'\i}a} {et~al.}(2011){Garc{\'\i}a}, {Hekker}, {Stello},
  {Guti{\'e}rrez-Soto}, {Handberg}, {Huber}, {Karoff}, {Uytterhoeven},
  {Appourchaux}, {Chaplin}, {Elsworth}, {Mathur}, {Ballot},
  {Christensen-Dalsgaard}, {Gilliland}, {Houdek}, {Jenkins}, {Kjeldsen},
  {McCauliff}, {Metcalfe}, {Middour}, {Molenda-Zakowicz}, {Monteiro}, {Smith},
  \& {Thompson}}]{Garcia11}
{Garc{\'\i}a}, R.~A., {Hekker}, S., {Stello}, D., {et~al.} 2011, \mnras, 414,
  L6

\bibitem[{{Garc{\'\i}a} {et~al.}(2014{\natexlab{b}}){Garc{\'\i}a}, {Mathur},
  {Pires}, {R{\'e}gulo}, {Bellamy}, {Pall{\'e}}, {Ballot}, {Barcel{\'o}
  Forteza}, {Beck}, {Bedding}, {Ceillier}, {Roca Cort{\'e}s}, {Salabert}, \&
  {Stello}}]{Garcia14}
{Garc{\'\i}a}, R.~A., {Mathur}, S., {Pires}, S., {et~al.} 2014{\natexlab{b}},
  \aap, 568, A10

\bibitem[{{Garc{\'{\i}}a} {et~al.}(2010){Garc{\'{\i}}a}, {Mathur}, {Salabert},
  {Ballot}, {R{\'e}gulo}, {Metcalfe}, \& {Baglin}}]{Garcia10Science}
{Garc{\'{\i}}a}, R.~A., {Mathur}, S., {Salabert}, D., {et~al.} 2010, Science,
  329, 1032

\bibitem[{{Gordon} {et~al.}(2021){Gordon}, {Davenport}, {Angus},
  {Foreman-Mackey}, {Agol}, {Covey}, {Ag{\"u}eros}, \& {Kipping}}]{Gordon21}
{Gordon}, T.~A., {Davenport}, J. R.~A., {Angus}, R., {et~al.} 2021, \apj, 913,
  70

\bibitem[{{Henden} {et~al.}(2015){Henden}, {Levine}, {Terrell}, \&
  {Welch}}]{Henden15}
{Henden}, A.~A., {Levine}, S., {Terrell}, D., \& {Welch}, D.~L. 2015, in
  American Astronomical Society Meeting Abstracts, Vol. 225, American
  Astronomical Society Meeting Abstracts \#225, 336.16

\bibitem[{{H{\o}g} {et~al.}(2000){H{\o}g}, {Fabricius}, {Makarov}, {Urban},
  {Corbin}, {Wycoff}, {Bastian}, {Schwekendiek}, \& {Wicenec}}]{Hog2000}
{H{\o}g}, E., {Fabricius}, C., {Makarov}, V.~V., {et~al.} 2000, \aap, 355, L27

\bibitem[{{Howell} {et~al.}(2014){Howell}, {Sobeck}, {Haas}, {Still},
  {Barclay}, {Mullally}, {Troeltzsch}, {Aigrain}, {Bryson}, {Caldwell},
  {Chaplin}, {Cochran}, {Huber}, {Marcy}, {Miglio}, {Najita}, {Smith},
  {Twicken}, \& {Fortney}}]{Howell14K2}
{Howell}, S.~B., {Sobeck}, C., {Haas}, M., {et~al.} 2014, \pasp, 126, 398

\bibitem[{Jeffreys(1961)}]{Jeffreys61}
Jeffreys, H. 1961, Theory of Probability (3rd Ed. OUP Oxford)

\bibitem[{{Koch} {et~al.}(2010){Koch}, {Borucki}, {Basri}, {Batalha}, {Brown},
  {Caldwell}, {Christensen-Dalsgaard}, {Cochran}, {DeVore}, {Dunham},
  {Gautier}, {Geary}, {Gilliland}, {Gould}, {Jenkins}, {Kondo}, {Latham},
  {Lissauer}, {Marcy}, {Monet}, {Sasselov}, {Boss}, {Brownlee}, {Caldwell},
  {Dupree}, {Howell}, {Kjeldsen}, {Meibom}, {Morrison}, {Owen}, {Reitsema},
  {Tarter}, {Bryson}, {Dotson}, {Gazis}, {Haas}, {Kolodziejczak}, {Rowe}, {Van
  Cleve}, {Allen}, {Chandrasekaran}, {Clarke}, {Li}, {Quintana}, {Tenenbaum},
  {Twicken}, \& {Wu}}]{Koch10}
{Koch}, D.~G., {Borucki}, W.~J., {Basri}, G., {et~al.} 2010, \apjl, 713, L79

\bibitem[{{Landin} {et~al.}(2010){Landin}, {Mendes}, \& {Vaz}}]{landin2010}
{Landin}, N.~R., {Mendes}, L.~T.~S., \& {Vaz}, L.~P.~R. 2010, \aap, 510, A46

\bibitem[{{Lehtinen} {et~al.}(2021){Lehtinen}, {K{\"a}pyl{\"a}}, {Olspert}, \&
  {Spada}}]{Lehtinen21}
{Lehtinen}, J.~J., {K{\"a}pyl{\"a}}, M.~J., {Olspert}, N., \& {Spada}, F. 2021,
  \apj, 910, 110

\bibitem[{{Lund} {et~al.}(2017){Lund}, {Silva Aguirre}, {Davies}, {Chaplin},
  {Christensen-Dalsgaard}, {Houdek}, {White}, {Bedding}, {Ball}, {Huber},
  {Antia}, {Lebreton}, {Latham}, {Handberg}, {Verma}, {Basu}, {Casagrande},
  {Justesen}, {Kjeldsen}, \& {Mosumgaard}}]{Lund17LEGACY}
{Lund}, M.~N., {Silva Aguirre}, V., {Davies}, G.~R., {et~al.} 2017, \apj, 835,
  172

\bibitem[{{Mathur} {et~al.}(2014{\natexlab{a}}){Mathur}, {Garc{\'\i}a},
  {Ballot}, {Ceillier}, {Salabert}, {Metcalfe}, {R{\'e}gulo}, {Jim{\'e}nez}, \&
  {Bloemen}}]{Mathur14b}
{Mathur}, S., {Garc{\'\i}a}, R.~A., {Ballot}, J., {et~al.} 2014{\natexlab{a}},
  \aap, 562, A124

\bibitem[{{Mathur} {et~al.}(2019){Mathur}, {Garc{\'\i}a}, {Bugnet}, {Santos},
  {Santiago}, \& {Beck}}]{Mathur19activity}
{Mathur}, S., {Garc{\'\i}a}, R.~A., {Bugnet}, L., {et~al.} 2019, Frontiers in
  Astronomy and Space Sciences, 6, 46

\bibitem[{{Mathur} {et~al.}(2010){Mathur}, {Garc{\'\i}a}, {R{\'e}gulo},
  {Creevey}, {Ballot}, {Salabert}, {Arentoft}, {Quirion}, {Chaplin}, \&
  {Kjeldsen}}]{Mathur10}
{Mathur}, S., {Garc{\'\i}a}, R.~A., {R{\'e}gulo}, C., {et~al.} 2010, \aap, 511,
  A46

\bibitem[{{Mathur} {et~al.}(2014{\natexlab{b}}){Mathur}, {Salabert},
  {Garc{\'\i}a}, \& {Ceillier}}]{Mathur14}
{Mathur}, S., {Salabert}, D., {Garc{\'\i}a}, R.~A., \& {Ceillier}, T.
  2014{\natexlab{b}}, Journal of Space Weather and Space Climate, 4, A15

\bibitem[{{McQuillan} {et~al.}(2014){McQuillan}, {Mazeh}, \&
  {Aigrain}}]{McQuillan14}
{McQuillan}, A., {Mazeh}, T., \& {Aigrain}, S. 2014, \apjs, 211, 24

\bibitem[{{Nielsen} {et~al.}(2013){Nielsen}, {Gizon}, {Schunker}, \&
  {Karoff}}]{Nielsen13}
{Nielsen}, M.~B., {Gizon}, L., {Schunker}, H., \& {Karoff}, C. 2013, \aap, 557,
  L10

\bibitem[{{Nissen} {et~al.}(2017){Nissen}, {Silva Aguirre},
  {Christensen-Dalsgaard}, {Collet}, {Grundahl}, \&
  {Slumstrup}}]{Nissen17LEGACY}
{Nissen}, P.~E., {Silva Aguirre}, V., {Christensen-Dalsgaard}, J., {et~al.}
  2017, \aap, 608, A112

\bibitem[{{Noyes} {et~al.}(1984){Noyes}, {Hartmann}, {Baliunas}, {Duncan}, \&
  {Vaughan}}]{Noyes84Rossby}
{Noyes}, R.~W., {Hartmann}, L.~W., {Baliunas}, S.~L., {Duncan}, D.~K., \&
  {Vaughan}, A.~H. 1984, \apj, 279, 763

\bibitem[{{Pires} {et~al.}(2015){Pires}, {Mathur}, {Garc{\'\i}a}, {Ballot},
  {Stello}, \& {Sato}}]{Pires15}
{Pires}, S., {Mathur}, S., {Garc{\'\i}a}, R.~A., {et~al.} 2015, \aap, 574, A18

\bibitem[{{Rauer} {et~al.}(2014){Rauer}, {Catala}, {Aerts}, {Appourchaux},
  {Benz}, {Brandeker}, {Christensen-Dalsgaard}, {Deleuil}, {Gizon}, {Goupil},
  {G{\"u}del}, {Janot-Pacheco}, {Mas-Hesse}, {Pagano}, {Piotto}, {Pollacco},
  {Santos}, {Smith}, {Su{\'a}rez}, {Szab{\'o}}, {Udry}, {Adibekyan}, {Alibert},
  {Almenara}, {Amaro-Seoane}, {Eiff}, {Asplund}, {Antonello}, {Barnes},
  {Baudin}, {Belkacem}, {Bergemann}, {Bihain}, {Birch}, {Bonfils}, {Boisse},
  {Bonomo}, {Borsa}, {Brand{\~a}o}, {Brocato}, {Brun}, {Burleigh}, {Burston},
  {Cabrera}, {Cassisi}, {Chaplin}, {Charpinet}, {Chiappini}, {Church},
  {Csizmadia}, {Cunha}, {Damasso}, {Davies}, {Deeg}, {D{\'{\i}}az}, {Dreizler},
  {Dreyer}, {Eggenberger}, {Ehrenreich}, {Eigm{\"u}ller}, {Erikson}, {Farmer},
  {Feltzing}, {de Oliveira Fialho}, {Figueira}, {Forveille}, {Fridlund},
  {Garc{\'{\i}}a}, {Giommi}, {Giuffrida}, {Godolt}, {Gomes da Silva},
  {Granzer}, {Grenfell}, {Grotsch-Noels}, {G{\"u}nther}, {Haswell}, {Hatzes},
  {H{\'e}brard}, {Hekker}, {Helled}, {Heng}, {Jenkins}, {Johansen},
  {Khodachenko}, {Kislyakova}, {Kley}, {Kolb}, {Krivova}, {Kupka}, {Lammer},
  {Lanza}, {Lebreton}, {Magrin}, {Marcos-Arenal}, {Marrese}, {Marques},
  {Martins}, {Mathis}, {Mathur}, {Messina}, {Miglio}, {Montalban}, {Montalto},
  {Monteiro}, {Moradi}, {Moravveji}, {Mordasini}, {Morel}, {Mortier},
  {Nascimbeni}, {Nelson}, {Nielsen}, {Noack}, {Norton}, {Ofir}, {Oshagh},
  {Ouazzani}, {P{\'a}pics}, {Parro}, {Petit}, {Plez}, {Poretti}, {Quirrenbach},
  {Ragazzoni}, {Raimondo}, {Rainer}, {Reese}, {Redmer}, {Reffert},
  {Rojas-Ayala}, {Roxburgh}, {Salmon}, {Santerne}, {Schneider}, {Schou},
  {Schuh}, {Schunker}, {Silva-Valio}, {Silvotti}, {Skillen}, {Snellen}, {Sohl},
  {Sousa}, {Sozzetti}, {Stello}, {Strassmeier}, {{\v S}vanda}, {Szab{\'o}},
  {Tkachenko}, {Valencia}, {Van Grootel}, {Vauclair}, {Ventura}, {Wagner},
  {Walton}, {Weingrill}, {Werner}, {Wheatley}, \& {Zwintz}}]{Rauer14PLATO}
{Rauer}, H., {Catala}, C., {Aerts}, C., {et~al.} 2014, Experimental Astronomy,
  38, 249

\bibitem[{{Reinhold} {et~al.}(2013){Reinhold}, {Reiners}, \&
  {Basri}}]{Reinhold13}
{Reinhold}, T., {Reiners}, A., \& {Basri}, G. 2013, \aap, 560, A4

\bibitem[{{Ricker} {et~al.}(2014){Ricker}, {Winn}, {Vanderspek}, {Latham},
  {Bakos}, {Bean}, {Berta-Thompson}, {Brown}, {Buchhave}, {Butler}, {Butler},
  {Chaplin}, {Charbonneau}, {Christensen-Dalsgaard}, {Clampin}, {Deming},
  {Doty}, {De Lee}, {Dressing}, {Dunham}, {Endl}, {Fressin}, {Ge}, {Henning},
  {Holman}, {Howard}, {Ida}, {Jenkins}, {Jernigan}, {Johnson}, {Kaltenegger},
  {Kawai}, {Kjeldsen}, {Laughlin}, {Levine}, {Lin}, {Lissauer}, {MacQueen},
  {Marcy}, {McCullough}, {Morton}, {Narita}, {Paegert}, {Palle}, {Pepe},
  {Pepper}, {Quirrenbach}, {Rinehart}, {Sasselov}, {Sato}, {Seager},
  {Sozzetti}, {Stassun}, {Sullivan}, {Szentgyorgyi}, {Torres}, {Udry}, \&
  {Villasenor}}]{Ricker14TESS}
{Ricker}, G.~R., {Winn}, J.~N., {Vanderspek}, R., {et~al.} 2014, in Society of
  Photo-Optical Instrumentation Engineers (SPIE) Conference Series, Vol. 9143,
  Society of Photo-Optical Instrumentation Engineers (SPIE) Conference Series,
  20

\bibitem[{{Ruediger} \& {Kichatinov}(1993)}]{1993A&A...269..581R}
{Ruediger}, G. \& {Kichatinov}, L.~L. 1993, \aap, 269, 581

\bibitem[{{Salabert} {et~al.}(2016){Salabert}, {Garc{\'\i}a}, {Beck},
  {Egeland}, {Pall{\'e}}, {Mathur}, {Metcalfe}, {do Nascimento}, {Ceillier},
  {Andersen}, \& {Trivi{\~n}o Hage}}]{Salabert16Sph}
{Salabert}, D., {Garc{\'\i}a}, R.~A., {Beck}, P.~G., {et~al.} 2016, \aap, 596,
  A31

\bibitem[{{Salabert} {et~al.}(2017){Salabert}, {Garc{\'{\i}}a}, {Jim{\'e}nez},
  {Bertello}, {Corsaro}, \& {Pall{\'e}}}]{Salabert17Sun}
{Salabert}, D., {Garc{\'{\i}}a}, R.~A., {Jim{\'e}nez}, A., {et~al.} 2017, \aap,
  608, A87

\bibitem[{{Santos} {et~al.}(2021){Santos}, {Breton}, {Mathur}, \&
  {Garc{\'\i}a}}]{Santos21}
{Santos}, A.~R.~G., {Breton}, S.~N., {Mathur}, S., \& {Garc{\'\i}a}, R.~A.
  2021, arXiv e-prints, arXiv:2107.02217

\bibitem[{{Santos} {et~al.}(2019){Santos}, {Garc{\'\i}a}, {Mathur}, {Bugnet},
  {van Saders}, {Metcalfe}, {Simonian}, \& {Pinsonneault}}]{Santos19}
{Santos}, A.~R.~G., {Garc{\'\i}a}, R.~A., {Mathur}, S., {et~al.} 2019, \apjs,
  244, 21

\bibitem[{{See} {et~al.}(2021){See}, {Roquette}, {Amard}, \&
  {Matt}}]{See21Rossby}
{See}, V., {Roquette}, J., {Amard}, L., \& {Matt}, S.~P. 2021, \apj, 912, 127

\bibitem[{{Silva Aguirre} {et~al.}(2017){Silva Aguirre}, {Lund}, {Antia},
  {Ball}, {Basu}, {Christensen-Dalsgaard}, {Lebreton}, {Reese}, {Verma},
  {Casagrande}, {Justesen}, {Mosumgaard}, {Chaplin}, {Bedding}, {Davies},
  {Handberg}, {Houdek}, {Huber}, {Kjeldsen}, {Latham}, {White}, {Coelho},
  {Miglio}, \& {Rendle}}]{SilvaAguirre17}
{Silva Aguirre}, V., {Lund}, M.~N., {Antia}, H.~M., {et~al.} 2017, \apj, 835,
  173

\bibitem[{Skilling(2004)}]{Skilling04}
Skilling, J. 2004, AIP Conference Proceedings, 735, 395

\end{thebibliography}
\end{document}